\documentclass[conference]{IEEEtran}
\IEEEoverridecommandlockouts
\usepackage{cite}
\usepackage{amsmath,amssymb,amsfonts}
\usepackage{algorithmic}
\usepackage{caption}
\usepackage{subcaption}
\usepackage{graphicx}
\usepackage{textcomp}
\usepackage{multirow}
\usepackage{xcolor}
\usepackage[margin=0.728in]{geometry}

\def\BibTeX{{\rm B\kern-.05em{\sc i\kern-.025em b}\kern-.08em
    T\kern-.1667em\lower.7ex\hbox{E}\kern-.125emX}}
\begin{document}

\title{Battery-Free Sensor Array for \\ Wireless Multi-Depth In-Situ Sensing\\

\thanks{This work was supported in part by NSF under grant no. CNS-2341846 and CNS-2310856.}
}

\author{\IEEEauthorblockN{ Hongzhi Guo and Adam Kamrath }
\IEEEauthorblockA{School of Computing, University of Nebraska-Lincoln, Lincoln, NE, USA \\
Email: hguo10@unl.edu; akamrath2@huskers.unl.edu}
}

\maketitle

\begin{abstract}
Underground in-situ sensing plays a vital role in precision agriculture and infrastructure monitoring. While existing sensing systems utilize wires to connect an array of sensors at various depths for spatial-temporal data collection, wireless underground sensor networks offer a cable-free alternative. However, these wireless sensors are typically battery-powered, necessitating periodic recharging or replacement. This paper proposes a battery-free sensor array which can be used for wireless multi-depth in-situ sensing. Utilizing Near Field Communication (NFC)---which can penetrate soil with negligible signal power loss---this sensor array can form a virtual magnetic waveguide, achieving long communication ranges. An analytical model has been developed to offer insights and determine optimal design parameters. Moreover, a prototype, constructed using off-the-shelf NFC sensors, was tested to validate the proposed concept. While this system is primarily designed for underground applications, it holds potential for other multi-depth in-situ sensing scenarios, including underwater environments.
\end{abstract}


\section{Introduction}
Multi-depth in-situ sensing systems are extensively used in agriculture and water management \cite{pichorim2018two,boada2018battery}. In precision agriculture, soil sensors are placed at various depths to comprehensively monitor attributes like salinity, moisture, and nitrogen density and distribution. This sensing data aids in the intelligent management of irrigation and nitrogen usage. Unlike single sensors, multi-depth sensory arrays gather rich information, providing a deeper understanding of a crop’s root nutrition.

\begin{figure}
     \centering
     \begin{subfigure}[b]{0.16\textwidth}
         \centering
         \includegraphics[width=\textwidth,height = 3.4cm]{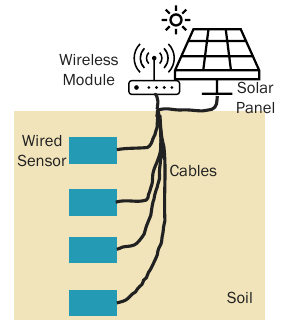}
         \caption{Wired sensor array using cables.}
         \label{fig:cable}
     \end{subfigure}
     \begin{subfigure}[b]{0.15\textwidth}
         \centering
         \includegraphics[width=\textwidth,height = 3.4cm]{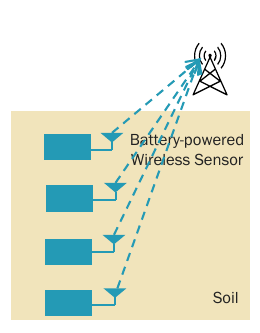}
         \caption{Battery-powered wireless sensor array.}
         \label{fig:wireless}
     \end{subfigure}
     \begin{subfigure}[b]{0.15\textwidth}
         \centering
         \includegraphics[width=\textwidth,height = 3.4cm]{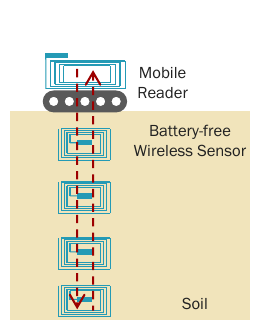}
         \caption{Battery-free wireless sensor array.}
         \label{fig:battery-free}
     \end{subfigure}
        \caption{Multi-depth in-situ sensory array.}
             \vspace{-10pt}
        \label{fig:sensor-array}
\end{figure}

Current multi-depth sensors are typically connected by cables, as shown in Fig.~\ref{fig:cable}. For easier deployment, these sensors can be housed in tubes. Often, a solar panel is positioned on the ground as the primary energy source. Additionally, the sensing system includes a wireless module on the surface that employs cellular communication or long-range Internet of Things (IoT) technologies to transmit the collected data. However, this system presents several drawbacks that have driven the search for alternative solutions. First, this type of sensing system is intrusive, which can alter the original distribution of the sensing parameters, leading to biased measurements. Second, the system's cost, often exceeding \$1,000, restricts widespread deployment. The system also necessitates specialized protection for the sensors, cables, and above-ground equipment, especially in extreme weather conditions. Last,  the current systems often experience data loss due to unreliable solar power sources or wireless channels.

In-situ wireless sensors eliminate the need for cable connections, as shown in Fig.~\ref{fig:wireless}. They are notably smaller and less intrusive compared to their wired counterparts. Rather than relying on solar panels for power, these sensors utilize batteries. While battery-powered wireless sensors have notably simplified deployment and reduced maintenance costs, they also present both environmental and technical challenges. First, to prolong battery life, the sensing system often operates in a low-power mode. This might entail reducing transmission power, increasing the time between sensing periods, and adopting energy-efficient cross-layer protocols. However, such measures can adversely affect both communication and sensing performance. Second, replacing batteries becomes a significant challenge, especially when sensors are positioned in inaccessible or unseen environments. Although wireless charging presents a possible solution, its efficiency diminishes considerably when sensors are buried deeply. Last, when the sensors reach the end of their operational life, they must be retrieved to ensure their batteries are disposed of responsibly. Such challenges are non-existent in terrestrial environments where locating and replacing or recharging batteries can be accomplished with relative ease. 

Battery-Free Wireless (BFW) sensors can harvest energy from electromagnetic fields. Once powered up, they can perform sensing tasks and transmit data by modulating signals over existing electromagnetic fields. Depending on the source of the electromagnetic fields, BFW sensors can either utilize ambient wireless signals or rely on dedicated signal generators. Due to their diminutive size, BFW sensors are non-intrusive; their presence in the soil does not significantly alter soil nutrition and water distribution. Ambient backscatter communication leverages legacy wireless sources such as LTE, Wi-Fi, and LoRa. While ambient backscatter communication is prevalent in terrestrial environments, legacy wireless signals struggle to penetrate soil or water, rendering them unsuitable for BFW sensors. Therefore, employing a dedicated wireless signal generator (often referred to as a "reader") becomes preferable, as it can provide robust transmission power. Current BFW sensors paired with a dedicated reader typically operate based on UHF (300 MHz to 3 GHz) RFID or HF (3 MHz to 30 MHz) RFID/NFC\cite{guo2021internet}. Given that soil possesses a certain water content and electrical conductivity, UHF signals face challenges penetrating deep soil --- for instance, the skin depth of soil at UHF frequencies is approximately 20 cm \cite{ellerbruch1974electromagnetic}. For deeper in-situ BFW sensing, HF RFID/NFC can be used due to their long wavelength. 

In this paper, we use HF NFC-based BFW sensors for wireless multi-depth in-situ sensing. An array of these sensors is evenly spaced and embedded within the soil, as shown in Fig.~\ref{fig:battery-free}. Above ground, a mobile reader initiates communication with the buried BFW sensors. These sensors harvest energy from the magnetic fields generated by the reader, then proceed to sense and respond in accordance with NFC anti-collision protocols. 

Deploying BFW sensors presents unique challenges, distinguishing them from their wired counterparts. First, many NFC modules possess a limited communication range, often less than 10 cm. Unlike battery-powered sensors that are inherently powered, battery-free sensors rely on downlink signals (from reader to sensor) for activation. Once powered, they can then remodulate these signals, transmitting them through uplink channels (from sensor to reader) \cite{ofori2022magnetic}. However, multi-depth in-situ sensing requires considerably longer communication ranges, spanning from 0.5 m to several meters, depending on the specific application. Thus, it's imperative to extend the communication range of NFC BFW sensors to align with these demands. Second, the HF band signals have a long wavelength and the coupling between sensor antennas must be carefully considered to avoid negative impacts on communication performance. Last, a near-far issue arises: sensors located closer to the reader receive significantly higher voltages than those situated further away. This poses a risk, as BFW sensors are low-power devices and excessive voltages could cause damage.   

To address theses challenges, this paper adopt the concept of magnetic induction waveguide \cite{sun2010magnetic,guo2015text}. In this approach, sensor coil antennas are spaced at regular intervals, forming a virtual link via magnetic coupling. While previous research has indicated that the magnetic induction waveguide can notably enhance communication range \cite{sun2010magnetic,guo2015text}, these studies predominantly focus on battery-powered sensors, leaving battery-free sensors unexplored. To our knowledge, this is the the first paper using magnetic induction waveguide concept for BFW sensors. The main contribution of this paper is two-fold. 
\begin{itemize}
    \item     We introduce an analytical communication and networking model for BFW sensors. Drawing from this, we offer insights into the optimal configurations and strategic placements of BFW sensors aimed at achieving a long communication range.
    \item We construct a testbed using readily available BFW sensors, testing our proposed approach both empirically and numerically. A comprehensive analysis of the empirical results, rooted in our analytical model, is also provided.
\end{itemize}

The rest of this paper is organized as follows. In Section II, we provide analytical communication models for a single BFW sensor and multiple BFW sensors and show the major difference. Then, we optimize the analytical model and obtain the optimal interval between sensors. The development testbed and numerical analysis are presented in Section III. Finally, this paper is concluded in Section IV.

\section{Communication System Modeling and Optimization}

NFC operates using coils that are coupled by near-field magnetic fields. While the coil antenna does produce electromagnetic waves, in the near field (where the distance from the antenna is much smaller than one wavelength) the electric and magnetic fields are weakly coupled. Unlike electromagnetic waves, near-field magnetic fields can penetrate non-magnetic materials with minimal scattering, reflection, and absorption losses. Consequently, NFC offers a reliable wireless channel, finding applications in diverse non-terrestrial environments \cite{zhao2020nfc+,guo2021internet,guo2021joint}. Communication and networking of battery-powered sensors using magnetic induction have been a subject of extensive research \cite{sun2010magnetic,kisseleff2016magnetic}. These can achieve extended communication ranges, exceeding 10 m, through specialized coil designs and circuit resonance. In contrast, BFW sensors have predominantly been employed for point-to-point communications rather than networked setups, owing to their inherently shorter communication ranges.
 
\begin{figure}[t]
    \centering
        \includegraphics[width=0.44\textwidth]{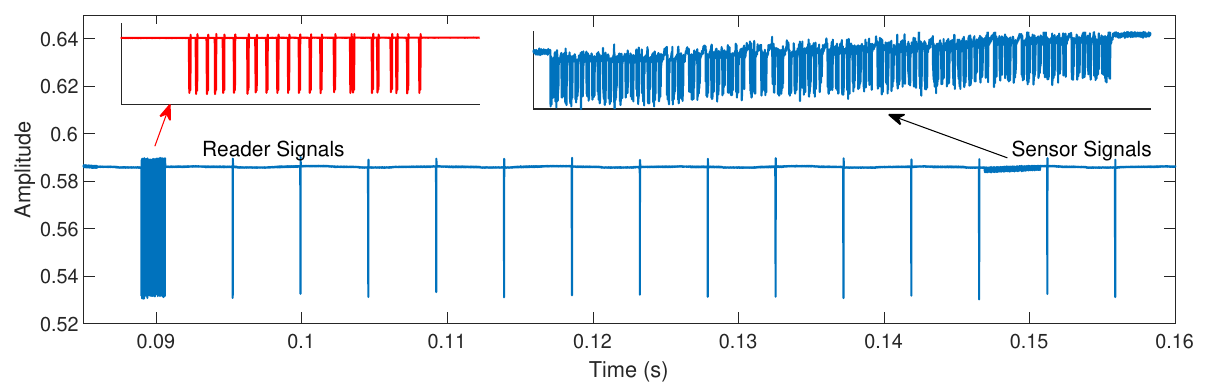}
        \vspace{-5pt}
        \caption{Illustration of ISO/IEC 15693 reader and sensor signals.}
    \label{fig:signal}
    \vspace{-10pt}
\end{figure}

NFC-based BFW sensors predominantly employ the ISO/IEC 14443 and ISO/IEC 15693 protocols \cite{guo2021internet}. Due to their reduced data rate, the ISO/IEC 15693 sensors can achieve a longer range compared to the ISO/IEC 14443 sensors. Ideally, with sufficiently large coil antennas, the ISO/IEC 15693 can support a communication range of up to 1.5 m. In this study, our BFW sensor array utilizes the ISO/IEC 15693 protocol. An illustration of the communication signal is provided in Fig.~\ref{fig:signal}. By positioning a USRP N210 equipped with an NFC antenna near a sensor, we captured the communication signals exchanged between the reader and the sensor. Following the reader's request transmission, it continues to emit continuous waves. Upon receipt of this request, the sensor selects an appropriate time slot for its response. As evident from the data, the signal from the sensor is considerably weaker than that from the reader.

\subsection{Joint Antenna and Channel Model}
\begin{figure}
     \centering
     \begin{subfigure}[b]{0.23\textwidth}
         \centering
         \includegraphics[width=\textwidth]{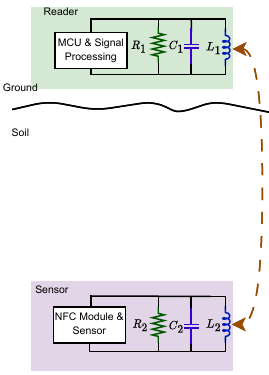}
         \vspace{6pt}
         \caption{Single sensor.}
         \label{fig:singlesensor}
     \end{subfigure}
     \begin{subfigure}[b]{0.23\textwidth}
         \centering
         \includegraphics[width=\textwidth]{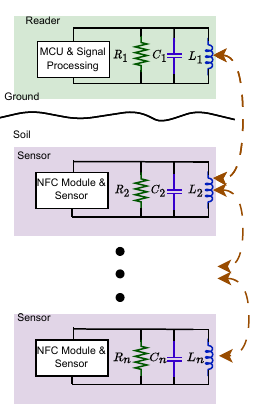}
         \caption{Multiple sensors.}
         \label{fig:multisensory}
     \end{subfigure}
        \caption{Illustration of the equivalent circuit for single sensor and multiple sensors.}
             \vspace{-10pt}
        \label{fig:circuit}
\end{figure}

Magnetic coupling between sensor coils can profoundly influence both the communication range and achievable data rates. Consequently, there's a need to develop a joint model for the antenna and channel. As depicted in Fig.~\ref{fig:circuit}, we utilize an equivalent parallel circuit model for both the reader and the sensors. This parallel circuit design is particularly effective at amplifying the induced voltage in the sensor circuits, ensuring it meets the cold start voltage requirements. This makes it a popular choice in NFC circuit design. In this paper, the reader's circuit elements and parameters are denoted using subscript 1, while the sensors' are denoted using subscript 2 to $n$ based on sensors' depth. The $n$th sensor has the largest depth. The self-inductance of coil $p$ is $L_p = \mu \pi a_p n_p^2/2$, where $\mu$ is the permittivity, $a_p$ is the $p$th sensor's coil radius, and $n_p$ is the $p$th sensor's coil number of turns \cite{sun2010magnetic}. The capacitor $C_p$ is used to tune the circuit to resonant, and $C_p = 1/(\omega^2 L_p)$, where $\omega=2\pi f_c$ and $f_c$ is the signal carrier frequency, i.e., 13.56 MHz. The resistor $R_p$ is used to control the coil's quality factor and bandwidth. If the quality factor $Q_p$, then $R_p = \omega L_pQ_p$. 

NFC coils use magnetic coupling to transmit signals and the wireless channel is determined by the mutual inductance. The mutual inductance between the $i$th and $j$th sensor coils is \cite{sun2010magnetic}
\begin{align}
\label{equ:mij}
    M_{ij}=\frac{\mu \pi n_i n_j a_i^2 a_j^2}{2d^3},
\end{align}
where $d$ is the distance between the coils. We can neglect the dynamics of soil conductivity and permittivity provided that $d$ is much smaller than the wavelength. The above equation can not provide accurate approximation when $d$ is extremely small. In this paper, we use the following mutual inductance equation when $d$ is small \cite{conway2007inductance}, 
\begin{align}
\label{equ:accurateM}
\begin{array}{r}
M_{ij}=\mu_0 \pi n_i n_j a_i a_j \int_0^{\infty} J_0(s p) J_1\left(s a_i\right) J_1\left(s a_j\right) \\
\exp \left(-s\left|z_2-z_1\right|\right) d s .
\end{array}
\end{align}
where $p$ is the perpendicular distance separating the coil axes, $|z_2-z_1|$ is the vertical distance separating the coil planes, $J_n(\cdot)$ is the Bessel Function of the $n^{th}$ order, and $s$ is the variable of integration. This model can also capture the horizontal deviation of sensors. Note that, although Equ.~\eqref{equ:accurateM} is accurate and generic, it is challenging to obtain insightful results using it. In the following, we will develop analytical models using Equ.~\eqref{equ:mij} and verify the results in Section~\ref{sec:sim} using Equ.~\eqref{equ:accurateM}.

\subsection{Single Sensor}
In existing NFC-based battery-free sensing systems, the reader communicates with each individual sensor. When there is one sensor, the communication range is determined by both the uplink and downlink channel. First, the downlink channel must induce sufficient voltage in the sensor's coil to power up the sensor. Second, the uplink channel must provide sufficient signal strength for the reader to demodulate the load modulated signals. 

Given the transmission power $P_t$, the voltage across the reader's coil is $v_1 = \sqrt{2\omega P_t L_1Q_1}$. The induced voltage across the sensor load is 
\begin{align}
    v_2 = v_{R_2 || C_2}=\frac{-j\omega L_2 R_2}{R_2- j\omega L_2}\cdot i_2
\end{align}
where $i_2$ is the current in the sensor coil. The current $i_2$ can be written as
\begin{align}
\label{equ:i2}
    i_2 = \frac{-j\omega M_{12} i_1}{-\frac{j\omega L_2 Q_2}{Q_2-j}+j\omega L_2}=\frac{M_{12}(Q_2-j)i_1}{jL_2}.
\end{align}
According to Kirchhoff's Voltage Law, the current in reader's coil satisfy the following condition
\begin{align}
\label{equ:v1}
    j\omega L_1 i_1 +j\omega M_{12} i_2 = v_1.
\end{align}
By substituting $i_2$ in Equ.~\eqref{equ:v1}, we can obtain $i_1$, which is then used to obtain $i_2$ and $v_2$. Since the sensor can only be power up when $v_2$ is larger than a threshold voltage, we pay attention to the amplitude of $v_2$, which is
\begin{align}
    |v_2| = \left|\frac{M_{12}Q_2L_2 v_1}{jL_1L_2+M_{12}^2(Q_2-j)}\right|
\end{align}
By expanding the expressions of $L_i$ and $M_{12}$, we can obtain
\begin{align}
\label{equ:v2single}
    |v_2| = \left| \frac{n_2 a_1 a_2^2Q_2\sqrt{\omega \mu \pi a_1 P_t Q_1} }{jd^3+\frac{a_1^3a_2^3(Q_2-j)}{d^3}} \right|
\end{align}
As we can see from the above equation, when $d$ is small increasing coil radii cannot effectively increase $|v_2|$ due to the second term in the denominator, while when $d$ is large, we can increase reader and sensor's coil radii, transmission power, quality factors, and sensor coil's number of turns to meet the threshold voltage. Since we are interested in the maximum range, the second term in the denominator can be neglected. As a result, the maximum downlink range is 
\begin{align}
    d_{max}^{dl} = \left( \frac{n_2 a_1 a_2^2Q_2\sqrt{\omega \mu \pi a_1 P_t Q_1} }{v_{th}} \right)^{\frac{1}{3}},
\end{align}
where $v_{th}$ is the threshold voltage. 

The reader detects load-modulated signals through the reflected impedance from the sensor coil. As shown in Equ.~\eqref{equ:v1}, the impact of the sensor coil can be written as
\begin{align}
    \alpha_s  = \left|\frac{\omega M_{12} i_2}{\omega L_1 i_1} \right|=  \frac{a_1^3a_2^3\sqrt{Q_2^2+1}}{d^6}.
\end{align}
Note that, there is a self-interference issue in the reader's coil. In Equ.~\eqref{equ:v1}, the first term on the left-hand side is for transmitting continuous waves to power sensors, while the second term reflects the signal modulated by the sensor. Both are in the same reader's coil. Use a strong transmission power can reduce the impact of noise and increase $d_{max}^{dl}$, but it cannot change $\alpha_s$. Given a threshold ratio $\alpha_t$, the maximum uplink range is 
\begin{align}
    d_{max}^{ul} = \left(\frac{a_1^3a_2^3\sqrt{Q_2^2+1}}{\alpha_t}\right)^{1/6}
\end{align}
As a result, the maximum signal sensor communication range is 
\begin{align}
    d_{max} = \min(d_{max}^{dl},d_{max}^{ul} ).
\end{align}
We will evaluate $d_{max}$ and compare it with the multi-sensor solution in Section \ref{sec:sim}.

\subsection{Multiple Sensors}
When there are multiple sensors, they can mutually affect each other. The impact can be positive or negative depending on sensor locations. Next, we develop a generic analytical model to characterize the multi-sensor interaction and obtain the optimal deployment policy. Considering the impact of all sensor coils, the reader's coil circuit satisfies,
\begin{align}
\label{equ:v12}
    j\omega L_1 i_1 +\sum_{k=2}^{n}j\omega M_{1k}i_{k} = v_1.
\end{align} 
Similarly, the voltage and current in a coil $p$ ($p\in [2,n]$) can be written as
\begin{align}
\label{equ:pvi}
    \sum_{k=1}^{p-1}j\omega M_{kp}i_{k}+Z_pi_p+\sum_{k=p+1}^{n}j\omega M_{kp}i_{k}=0,
\end{align}
where $Z_p$ is the impedance of the coil $p$. 

\subsubsection{Approximated Analytical Model}
To obtain the induced voltage in each sensor coil, we can rewrite the above equations in matrix format, as used in \cite{kisseleff2016magnetic}. However, the challenge is not to obtain the voltage in sensor coils, but to optimize sensors' location to maximize the communication range. In this paper, we decompose this complex problem using the following two steps. 

First, the mutual inductance between nonadjacent coils is small. In the sensory array, only the top and the bottom (if any) sensor coils can strongly affect the sensor coil in the middle. As a result, Equ.~\eqref{equ:pvi} can be rewritten as
\begin{align}
\label{equ:simplify}
    j\omega M_{1p}i_{1}+j\omega M_{(p-1)p}i_{p-1}+Z_pi_p+j\omega M_{p(p+1)}i_{p+1}\approx 0.
\end{align}
In Equ.~\eqref{equ:simplify}, we keep the induced voltage by the reader coil. This is because in NFC systems, the reader coil usually has a larger size than sensor coils and $M_{1p}$ can be relatively large. 

Second, the sensors form a chain to relay signals and sensor coil currents on this chain should decrease at a constant rate, i.e, $i_{p-1}\beta= i_{p}$ for $p \in [2, n-1]$, where $|\beta| \leq 1$. Then, we search for a sensor deployment solution that can maximize $\beta$ given a minimum interval. Since the considered environment is homogeneous and sensors are placed with a constant interval, the optimal mutual inductance between two sensor coils should be similar. As a result, Equ.~\eqref{equ:simplify} can be rewritten as
\begin{align}
    \gamma j\omega M i_{p-1}+j\omega M i_{p-1}+Z_p i_{p-1}\beta + j\omega M i_{p-1} \beta^2=0,
\end{align}
where we consider $j\omega M_{1p}i_{1}=\gamma j\omega M i_{p-1}$. Then, we can find $\beta$, which is 
\begin{align}
    \beta = \frac{-Z_p+\sqrt{Z_p^2+4\omega^2 M^2(1+\gamma)}}{2j\omega M}.
\end{align}
Since the sensor coil coupling $\omega M$ is usually much smaller than $Z_p$, we can obtain the following approximation using Taylor Series,
\begin{align}
    \beta\approx \frac{\frac{1}{2}\frac{4\omega^2 M^2(1+\gamma)}{Z_p^2}}{\frac{2j\omega M}{Z_p}}=\frac{\omega M (1+\gamma)}{j Z_p}.
\end{align}
Since $Z_p$ can be expressed as $-j\omega L_pQ_p/(Q_p-j)+j\omega L_p$, $\beta$ can be simplified to
\begin{align}
\label{equ:beta}
    |\beta| \approx\frac{a_p^3}{d^3}(1+\gamma)\sqrt{Q_p^2+1}. 
\end{align}
Here, we implicitly assume that all the sensor coils have the same number of turns and the same coil radius. 

Next, we consider the reader's impact can be neglected, i.e., $\gamma=0$, and the coil's quality factor is much larger than 1, i.e., $Q_p>>1$. Given $\beta$, the distance between two sensors is 
\begin{align}
\label{equ:d}
    d = \sqrt[3]{\frac{Q_p}{|\beta|}}a_p.
\end{align}
As we can see from the above equation, we can increase sensor coil's quality factor and size to extend the range. Also, using a small $\beta$ can also increase the range between two sensors, however, that decreases the power delivered to the deeper sensor which can result in a short overall communication range. 

Since the sensor currents are proportional, we can find $i_2$ then based on $\beta$, we can obtain the rest of the currents. In Equ.~\eqref{equ:v12}, the summation can only include sensor 2 and neglect other sensor coils due to their weak currents and small mutual inductance. Then, by using Equ.~\eqref{equ:v12} and Equ.~\eqref{equ:pvi} (only consider adjacent mutual inductance), we can obtain
\begin{align}
\label{equ:i2}
    i_2= \frac{-M_{12}(Q_2-j)v_1}{-j\omega M_{12}^2(Q_2-j)+\omega L_1 L_2+j\omega M_{23}L_1\beta (Q_2-j)}. 
\end{align}
We can approximately obtain the current in each sensor coil which is $i_p \approx \beta^{p-2} i_2$. The voltage in each sensor is $|v_p|=|\omega M_{(p-1)p} i_{p-1}Q_p|$. Note that, we cannot get a simple equation for $d_{max}^{dl}$ when there are multiple sensors. The system can operate only if the sensor voltage is higher than $v_{th}$. 

From Equ.~\eqref{equ:i2} we can see that if sensor 2 and sensor 3 are extremely close, i.e., $M_{23}$ is large, $i_2$ can be small. As a result, sensor 2 cannot receive enough voltage to power up. This shows that when there are multiple sensors, their distance need to be carefully planned to avoid negative effects. 

The uplink for sensor $k$ in multi-sensor scenario can be evaluated using the $j\omega M_{1k}i_k$, which is 
\begin{align}
    \alpha_m^p = \left|\frac{\omega M_{1k}i_k}{\omega L_1 i_1+\sum_{p, p\neq k}^{n}\omega M_{1 p}i_p}\right|.
\end{align}
The reader can demodulate sensor data only if $\alpha_m^p>\alpha_t$. 

\subsubsection{Optimal Multi-Depth Sensor Deployment}
\label{sec:optimal}
In the previous part, we present the relation between distance and current decay rate. In multi-depth in-situ sensing, we are interested in several depths with the same mutual interval, e.g., the depth of sensor $p$ is $l_p = (p-1) l_0$, where $p \in[2,n]$ is the index of the sensor and $l_0$ is the interval. Given the required depths, we need to find the optimal or feasible combinations of sensor coil radius $a_p$, quality factor $Q_p$, and reader transmission power $P_t$ to achieve the range.

As we can see from Equ. \eqref{equ:beta} to Equ.~\eqref{equ:i2}, the coil radius and quality factor can be considered jointly and it is challenging to separate them. We use a search-based solution. We consider $a_p=a_0$ and $Q_p=Q_0$ which are standard values used by NFC sensors, e.g., $a_0=2.5$~cm and $Q_p=8$. The transmission power is $P_t=0.01$~W. We solve equations \eqref{equ:v12} and \eqref{equ:pvi} and check if the voltages can meet the threshold voltage requirement. If not, we update $Q_p=Q_p+\Delta Q$, $P_t = P_t+\Delta P_t$, and $a_p=a_p+\Delta a$ in sequence, and then check the voltages iteratively. This process will continue until we obtain the feasible $a_p$, $Q_p$, and $P_t$. Also, we set the maximum radius $a_p^{max}$, maximum quality factor $Q_p^{max}$, and maximum transmission power $P_t^{max}$. When all of them reach the maximum values but we cannot obtain a feasible solution, the requirement is beyond the capability of the system. 

\section{Implementation and Simulation}
\label{sec:sim}
In this section, we show an implementation using off-the-shelf NFC sensors and understand the results using the developed analytical models. Also, we provide more simulations to show the achievable communication range of the proposed system. 
\subsection{Implementation using Off-The-Shelf Sensors}
In our implementation, we consider the following off-the-shelf reader and sensors. 
\begin{itemize}
    \item NFC battery-free sensor. We use the MAX66242EVKIT sensor board from Analog Devices. The sensor board has a temperature sensor and supports ISO/IEC 15693 protocol.  
    \item Long-range reader. We use the FEIG ID MR102 reader with ANT340/240 pad antenna which supports ISO/IEC 15693 protocol. The reading range is around 35 cm.  
\end{itemize}

\begin{figure}[t]
    \centering
        \includegraphics[width=0.44\textwidth]{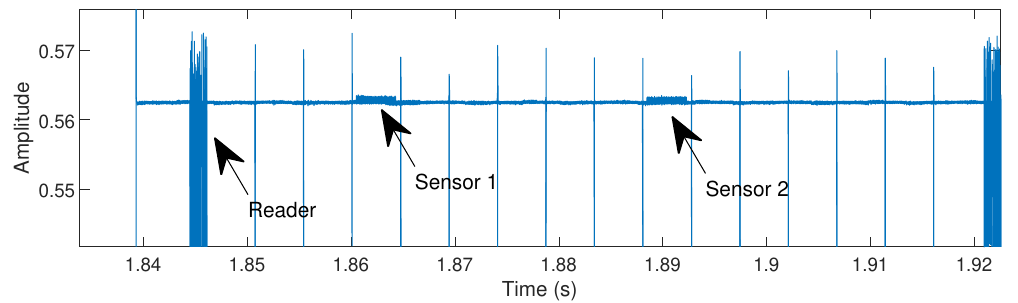}
        \caption{Signals with two sensors' responses.}
    \label{fig:signal2}
\end{figure}

\begin{figure}
     \centering
     \begin{subfigure}[b]{0.2\textwidth}
         \centering
         \includegraphics[width=\textwidth,height = 3.4cm]{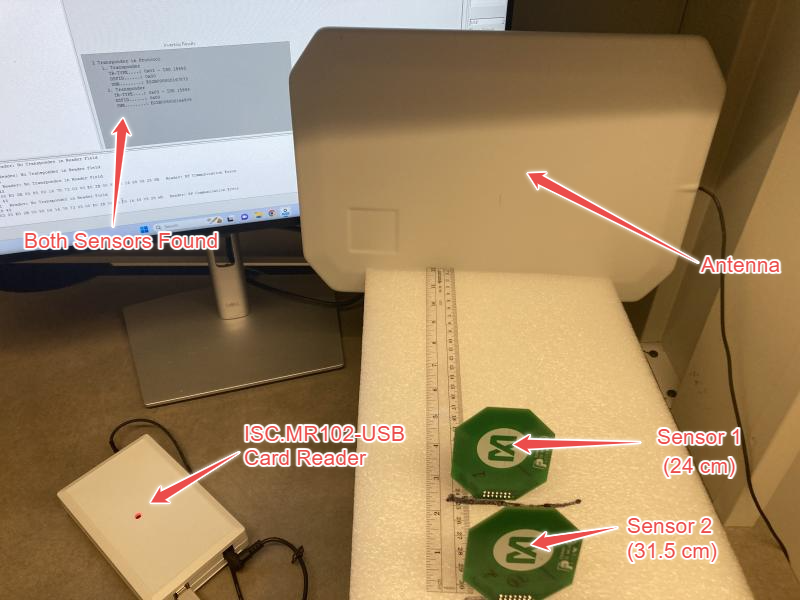}
         \caption{Two sensors in the air.}
         \label{fig:air}
     \end{subfigure}
     \begin{subfigure}[b]{0.2\textwidth}
         \centering
         \includegraphics[width=\textwidth,height = 3.4cm]{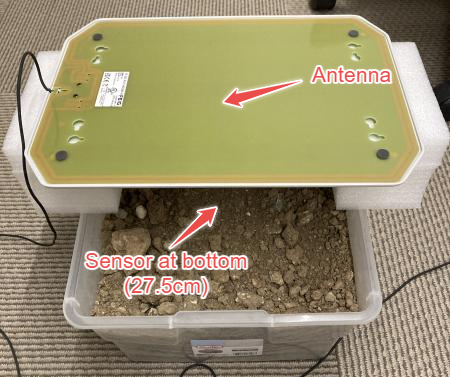}
         \caption{One sensor in a soil box.}
         \label{fig:soil}
     \end{subfigure}
        \caption{Measurement setup.}
             \vspace{-15pt}
        \label{fig:measurement}
\end{figure}

We measured the response signals from two sensors, as shown in Fig.~\ref{fig:signal2}. In order to obtain a clear signal, we place the two sensors close to the reader's antenna. As we can see, the sensors respond in different time slots to avoid collisions. We test the multi-depth sensor array in air and soil to understand the impact of soil. The measure setup is shown in Fig.~\ref{fig:measurement}. We measure the maximum communication range (where the reader can read a sensor's ID) in the air for one sensor and two sensors. Figure \ref{fig:air} shows the two-sensor setup. Then, we place a sensor in a soil box with soil depth of around 15 cm and vary the height of the reader's antenna, as shown in Fig.~\ref{fig:soil}.   

\begin{figure*}[htbp]
	\begin{minipage}{0.3\linewidth}
		\centering
		\includegraphics[width=0.99\textwidth,height=4.5cm]{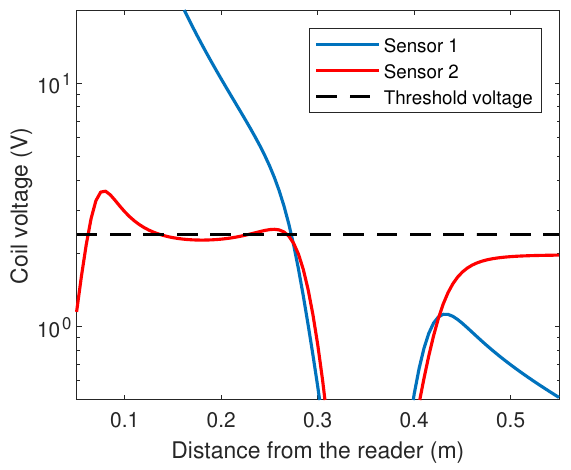}
		\caption{Induced voltages in a two-sensor case.}
		\label{fig:analytical_exp}
	\end{minipage}%
	\quad
	\begin{minipage}{0.3\linewidth}
		\centering
		\includegraphics[width=0.99\textwidth,height=4.5cm]{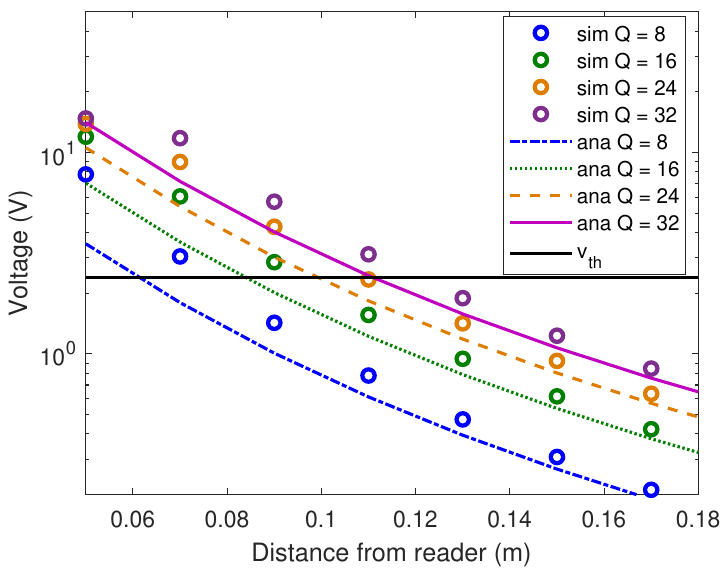}
		\caption{Downlink communication range between a reader and a sensor.}
		\label{fig:single}
	\end{minipage}
	\quad
	\begin{minipage}{0.295\linewidth}
		\centering
		\includegraphics[width=0.99\textwidth,height=4.5cm]{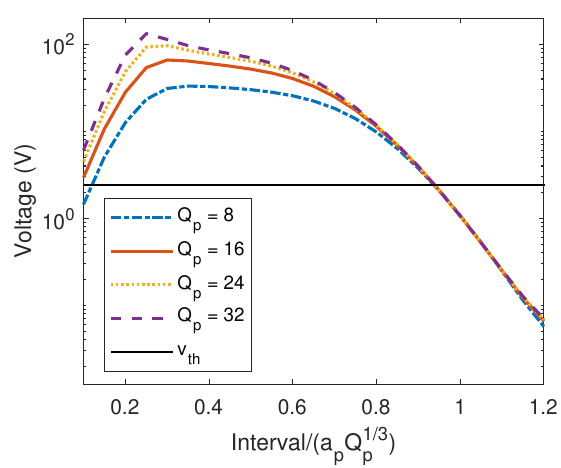}
		\caption{Impact of quality factor on optimal interval between two sensors.}
		\label{fig:interval}
	\end{minipage}
	\vspace{-15pt}
\end{figure*} 

As shown in Table~\ref{tab:measure}, for one sensor, the communication ranges in the air and in the soil are similar which shows that the soil does not have significant impact on the sensor coil and signal propagation. In the air, we use two sensors and form an array and the overall range can be extended. However, it only works when sensor 1 and sensor 2 are placed at specific locations. This is mainly due to the low quality factor of sensor coils and their small sizes. Note that, when sensor 1 is close to the reader, the simulation results show that the sensor 2 can be powered up. However, in experiment, we cannot observe this phenomenon. The sensor 1 cannot support high voltages (e.g., higher than 20 V). It is impractical to use this region. On the contrary, when sensor 1 is close to sensor 2 and the induced voltage in sensor 2 is slightly higher than the threshold voltage. Meanwhile, sensor 1's voltage is not high. As sensor 1 moves further closer, the coupling between two sensor coils becomes so strong that neither of them can communicate. In Fig.~\ref{fig:analytical_exp}, we choose parameters to approximate the experiment measures and show this phenomenon using the analytical model. As we can see in the figure, theoretically, there are two regions that sensor 2 can be powered up. However, the region that is close to the reader induces strong voltages in sensor 1. Sensor 1's circuit has protection mechanism which does not allow such strong voltages. Therefore, only the region close to sensor 2 can be measured in practice. 

As a result, there is a ``near-far'' problem here. When we try to power up a far sensor using high transmission power or high-quality-factor transmission coils, the near sensors will be exposed to high voltages which can destroy their circuits. However, the near sensors have to be alive to relay signals for far sensors. This issue can be addressed by using a transformer to separate the sensor coil and sensor circuits. The transformer needs to be specifically designed for each sensor with different depths. Due to the limitation of space, we do not consider transformer design in this paper.  
\begin{table}[hb]
    \caption{Range Measurement.}
    \label{tab:measure}
    \centering
\begin{tabular}{ |c|c|c|c| } 
\hline
 & Air (1 sensor) & Air ( 2 sensors) & Soil (1 sensor) \\ \hline
Sensor 1 & 29.5 cm & 24 cm & 27.5 cm\\ \hline
Sensor 2& --- & 31.5 cm &--- \\ 
\hline
\end{tabular}

\vspace{-13pt}
\end{table}

\subsection{Numerical Simulation}
In the following, we evaluate the impact of various sensor coil configurations on the communication range. First, we consider standard NFC reader and sensor coils. The reader coil radius is 4 cm and sensor coil radius is 2.5 cm. Both the reader and sensor coil have 5 turns. The reader coil quality factor is 8 and the sensor coil quality factor is varied from 8 to 32. The transmission power is 0.01 W.    

As shown in Fig.~\ref{fig:single}, the downlink communication range is determined by the power-up threshold voltage $v_{th}$. As the sensor coil quality factor increases from 8 to 32, the range is increased from 6 cm to around 11 cm. Due to the short communication range, we used Equ.~\eqref{equ:accurateM} to simulate (sim in Fig.~\ref{fig:single}) the induced voltage, and the analytical model (ana in Fig.~\ref{fig:single}) uses Equ.~\eqref{equ:v2single}. The results are consistent with existing NFC sensors. For example, the measured maximum communication range between MAX66300-24XEVKIT and MAX66242EVKIT is 7 cm. Although increasing the quality factor can extend the range, using existing NFC sensors cannot meet most long-range multi-depth in-situ sensing requirements. 


In Equ.~\eqref{equ:d}, we find that the optimal interval is proportional to $a_p\sqrt[3]{Q_p}$. In Fig.~\ref{fig:interval}, we show the induced voltage in the deepest sensor of a 10-sensor array. Their interval is scaled by $a_p\sqrt[3]{Q_p}$ and shown in the x-axis. We consider four sensor coil quality factors. Since the interval is scaled by $a_p\sqrt[3]{Q_p}$, the induced voltage in sensor coils should be similar. As we can see in Fig.~\ref{fig:interval}, as the interval increases sensor voltages converges to the same value. When the interval is small, the reader has stronger impact on each sensor coil. As a result, the assumption of neglecting reader impact to derive Equ.~\eqref{equ:d} is not valid when the distance from the reader is short. 


\begin{figure}[t]
    \centering
        \includegraphics[width=0.38\textwidth]{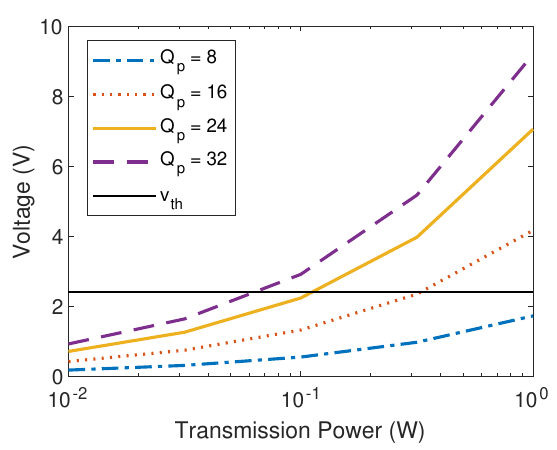}
        \vspace{-5pt}
        \caption{Required transmission power to achieve 0.6 m range using sensor arrays with various sensor coil quality factor.}
    \label{fig:power}
    \vspace{-10pt}
\end{figure}
In Fig.~\ref{fig:power}, we set the required range as 0.8 m and set the interval as $0.8a_p\sqrt[3]{Q_p}$. The interval and the overall range will determine the number of required sensors in the sensor array. The transmission power is varied from 0.01 W to 1 W. As we can see in Fig.~\ref{fig:power}, the high quality factor can effectively reduce the use transmission power by achieving the threshold voltage using a lower transmission power.  

\begin{figure}[t]
    \centering
        \includegraphics[width=0.38\textwidth]{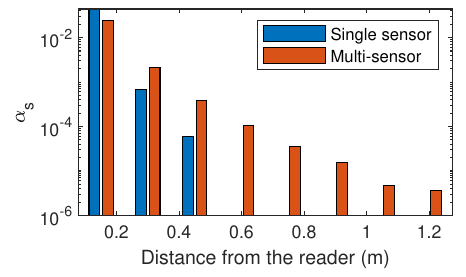}
        \vspace{-5pt}
        \caption{$\alpha_s$ of signal sensor at 8 different locations and $\alpha_s$ for 8 sensors in a multi-depth sensor array.}
        \vspace{-10pt}
    \label{fig:uplink}
\end{figure}

We also evaluated the optimal multi-depth sensor deployment. Given a requirement sensing depth of 1.2 m with 0.15 m interval, we use 8 sensors. By iterating the quality factor, transmission power, and sensor coil radius as discussed in Section~\ref{sec:optimal}, we find that we can successfully power up all the sensors using $Q_p=32$, $P_t=1$ W, and $a_p=$0.05 m. To evaluate the uplink channel, we use the derived optimal multi-depth sensor array and obtain the $\alpha_s$ for each sensor. The results are compared with a single sensor at various depths as the sensor array. As shown in Fig.~\ref{fig:uplink}, the single sensor can only receive enough voltage at the first three locations, while the multi-depth sensor array can receive enough voltage at 1.2 m away. $\alpha_s$ is decreasing as the distance increases.  

\section{Conclusion}
In this paper, we propose a battery-free multi-depth sensing system, where sensors use NFC modules for data communication. The sensor coils form a virtual magnetic induction waveguide to achieve a longer communication range than each individual sensor. We develop an analytical model to identify key parameters and design the optimal sensor coil parameters. A proof-of-concept testbed is developed and numerical simulations are used to comprehensively evaluate the performance. 

\bibliographystyle{IEEEtran}
\bibliography{ref}

\end{document}